\documentclass[useAMS]{mn2e}
\usepackage{times}
\usepackage{psfig}
\usepackage{epsfig}
\usepackage{graphicx}

\def\lsim{\lower.5ex\hbox{$\; \buildrel < \over \sim \;$}}
\def\gsim{\lower.5ex\hbox{$\; \buildrel > \over \sim \;$}}

\def\lsim{\lower.5ex\hbox{$\; \buildrel < \over \sim \;$}}
\def\gsim{\lower.5ex\hbox{$\; \buildrel > \over \sim \;$}}

\def\baselinestretch{1.}

\def\be{\begin{equation}}

\def\ee{\end{equation}}

\begin{document}

\title[Nonlinear electrodynamics and gravitational redshift of ...]{Nonlinear electrodynamics and the gravitational redshift of highly magnetised neutron 
stars}

\author[Mosquera Cuesta and Salim]{Herman J. Mosquera Cuesta$^{1,2,3}$ and 
Jos\'e M. Salim$^1$ \\
$^1$Instituto de Cosmologia, Relatividade e Astrof\'{\i}sica (ICRA-BR), 
Centro Brasileiro de Pesquisas F\'{\i}sicas \\ Rua Dr. Xavier Sigaud 150,
Cep 22290-180, Urca, Rio de Janeiro, RJ, Brazil \\  
$^2$Abdus Salam International Centre for Theoretical Physics\\ Strada 
Costiera 11, Miramare 34014, Trieste, Italy \\
$^3$Centro Latino-Americano de F\'{\i}sica, Avenida Wenceslau Braz 
71, CEP 22290-140 Fundos, Botafogo, Rio de Janeiro, RJ, Brazil }

\maketitle

\date{\today}

\begin{abstract}
We show that nonlinear electrodynamics (NLED) modifies in a fundamental 
basis the concept of gravitational red-shift (GRS) as it was introduced 
by Einstein's general relativity. The effect becomes apparent when light 
propagation from super strongly magnetized compact objects, as pulsars, 
is under focus. The analysis, here based on the (exact) nonlinear Lagrangean 
of Born \& Infeld (1934), proves that alike from general relativity (GR);  
where the GRS independs on any background magnetic ($B$) field, when NLED 
is taken into the photon dynamics an effective GRS appears, which happens 
to decidedly depend on the $B$-field pervading the pulsar. The resulting 
GRS tends  to infinity as the $B$-field grows larger, as opposed to the 
Einstein prediction. As in astrophysics the GRS is admittedly used to 
infer the mass-radius relation, and thus the equation of state of a 
compact star, e.g. a neutron star (Cottam, Paerels \& Mendez 2002), this 
unexpected GRS critical change may misguide observers to that fundamental 
property. Hence, a correct procedure to estimate those valuous physical 
properties demands a neat separation of the NLED effects from the pure 
gravitational ones in the light emitted by ultra magnetised pulsars.
\end{abstract}


\section{INTRODUCTION}

The idea that the nonlinear electromagnetic interaction, i. e., light
propagation in vacuum, can be geometrized was developed by Novello et
al. (2000) and Novello \& Salim (2001). Since then a number of physical
consequences for the dynamics of a variety of systems have been explored. 
In a recent paper, Mosquera Cuesta \& Salim (2003,2004) presented the first 
astrophysical context where such nonlinear electrodynamics effects were 
accounted for: the case of a highly magnetized neutron star or pulsar. 
In that paper NLED was invoked {\it a l\`a} Euler-Heisenberg, which is an 
infinite series expansion of which only the first nonlinear term was used 
for the analisys. An immediate consequence of that study was an overall 
modification of the space-time geometry around the pulsar as ``perceived''
by light propagating out of it. This translates into a fundamental change 
of the star surface redshift, the GRS, which might have been inferred from 
the absorption (or emission) lines observed in a super magnetized pulsar 
by Ibrahim et {al.} (2002;2003). The result proved to be even more dramatic 
for the so-called magnetars: pulsars said to be endowed with magnetic ($B$) 
fields higher then the Schafroth quantum electrodynamics critical $B$-field. 
In this {\it Letter} we demonstrate that the same effect still appears if 
one calls for the NLED in the form of the one rigorously derived by Born \& 
Infeld (1934), which is based on the special relativistic limit to the 
velocity of approaching of an elementary charged particle to a pointlike 
electron. As compared to our previous results, here we stress that from 
the mathematical point of view the Born \& Infeld (1934) NLED is described 
by an exact Lagrangean whose dynamics has been successfully studied in a 
wide set of physical systems (Delphenich 2003). The analysis presented next 
proves that this physics affects not only the magnetar electrodynamics (our 
focus here) but also that one of newly born; highly magnetised proto-neutron 
stars and the dynamics of their progenitor supernovae.

\section{Neutron star mass-radius relation}

Neutron stars (NSs), formed during the death throes of massive stars, are
among the most exotic objects in the universe. They are supposed to be
composed of essentially neutrons, although some protons and electrons
are also required in order to guarantee stability against Pauli's
exclusion principle for fermions. As remnants of supernova explosions
or accretion-induced-collapse of white dwarfs, they are (canonical)
objects with extremely high density $ \rho \sim 10^{14}$~g
cm$^{-3}$ for a mass $\sim 1~$M$_\odot$ and radius $R \sim 10$~km, and 
supposed to be endowed with typical magnetic fields of about 
$B \leq 10^{12}$~G, as expected from magnetic flux conservation during 
the supernova collapse.

In view of its density, a neutron star (NS) is also believed to trap 
in its core a substantial part of even more exotic states of matter. 
It is almost a concensus that these new states might exist inside and 
may dominate the star structural properties. Pion plus kaon Bose-Einstein
condensates could appear, as well as ``bags" of strange quark matter
(Miller 2002). This last one is believed to be the most stable state
of nuclear matter (Glendenning 1997), which implies an extremely
dense medium whose physics is currently under severe scrutiny.  The
major effect of these exotic constituents is manifested through the
NS mass-to-radius ratio ($M/R$). Most researchers in the field
think of the presence of such exotic components not only as to make the
star more compact, i.e., smaller in radius, but also to lower the maximum
mass it can retain.

The fundamental properties of NSs, their mass ($M$) and radius ($R$),
provide a direct test of the equation of state (EOS) of cold nuclear
matter, a relationship between pressure and density that is determined
by the physics of the strong interactions between the particles that
constitute the star. It is admitted that the most direct procedure of
estimating these properties is by measuring the gravitational redshift
of spectral lines produced in the NS photosphere. As the EOS relates
directly $M$ and $R$, hence a measurement of the GRS at the star
surface leads to a strong constraint on the $M/R$ ratio. In
this connection, observations of the low-mass X-ray binary EXO0748-676
by Cottam, Paerels and Mendez (2002) lead to the discovery of
absorption lines in the spectra of a handful of X-ray bursts, with most
of the features associated with Fe$^{26}$ and Fe$^{25}$ $n = 2-3$ and
O$^{8}$ $n =  1-2$ transitions, all at a redshift $z = 0.35$ (see its
definition in Eq.(\ref{redshift}) below). The conclusions regarding the
nature of the nuclear matter making up that star thus seem to exclude
some models in which the NS material is composed of more exotic matter
than the cold nuclear one, such as the strange quark matter or kaon
condensates for $M = 1.4-1.8$ ~M$_\odot$ and $R = 9-12$~km.

But the identification of absorption lines was also achieved by Sanwal
et al. (2002) using Chandra observations of the isolated NS 
1E1207.4-5209.  The lines observed were found to correspond to energies
of 700 and 1.400~eV, which they interpreted as the signature of singly
ionized helium in a strong magnetic field\footnote{Note that the $B$-field 
strength in this source is unknown because no spindown was measured in 
it.}. The inferred redshift is 0.12-0.23. Since, as well as gravity (see
Eq.(\ref{redshift})), a magnetic field has nontrivial effects on the
line energy, e.g. Eq.(\ref{line}), then Sanwal et al. (2002) could not
make their case for a correct and accurate identification of the lines,
and neither can they decidedly rule out alternative interpretations,
such as the cyclotron feature, which is expected from interacting 
X-ray binaries, as prescribed by the relation 

\be
E_{\rm He} = {3.2} \left(1 + z\right)^{-1} \left[\frac{B_{\rm sc}}{10^{15} 
\rm G}\right] \; {\rm keV}\label{line}\; .
\ee

To get some insight into the NS most elusive properties: its mass and
radius, astronomers use several techniques at disposal.  Its mass can
be estimated, in some cases, from the orbital dynamics of X-ray binary
systems, while attempts to measure its radius proceed via
high-resolution spectroscopy, as done very recently by Sanwal et al.
(2002) upon studying the star 1E1207.4-5209; and Cottam, Paerels and
Mendez (2002) by analysing type I X-ray bursts from the star
EXO0748-676. In those systems success was achieved in determining those
parameters, or the relation between them, by looking at the general
relativistic effect known as GRS of excited ions near the NS surface.
Gravity effects cause the observed energies of the spectral lines of
excited atoms to be shifted to lower values by a factor

\be
\frac{1}{(1 + z)}  \equiv \left( 1 - \frac{2 G}{c^2} \left[\frac{M}{R }
\right] \right)^{1/2} \; , \label{redshift}
\ee

with $z$ the GRS. Since this redshift depends on $M/R$, then measuring 
the spectral lines displacement leads to an indirect, but highly accurate, 
estimate of the star radius.

The above analysis stands on whenever the effects of the NS $B$-field
are negligible. However, if the NS  is pervaded by a super strong
$B$-field ($B_{\rm sc}$), as in the so-called {\it magnetars}, there is
then the possibility, for a given field strength, for the gravity
effects to be emulated by the electromagnetic ones. In what follows we
prove that this is the case if NLED [{\it a l\`a } Born-Infeld (1934)]
is taken into account to describe the general physics taking place on
the pulsar surface. Our major result proves that for very high magnetic 
fields ($B \leq 10^{14}$~G) the redshift induced by NLED can be as high 
as that one produced by gravity alone, while in the extreme limit ($B 
\geq 10^{15}~$G) it largely overtakes it (see Figs.1 and 2). This way the 
NLED emulates the gravitation. In such stars, then, care should be excersized 
when putting forward claims regarding the $M/R$ or the EOS of the observed 
pulsar.

\section{NLED and effective metric}

It is well-known that extremely strong magnetic fields induce the phenomenon
 of {\it vacuum polarization}, which manifests, depending on the magnetic 
field strength, as either real or virtual electron-positron pair creation 
in a vacuum. Although this is a quantum effect, we stress that it can also 
be described classically by including corrections to the standard linear
Maxwell's Lagrangean. Because of their interaction with this excited 
background of pairs, the modification of the {\it dispersion relation} 
for photons, as compared to the one from Maxwell dynamics, is one of the 
major consequences of the nonlinearities introduced by the NLED Lagrangean. 
Among the principal alterations associated to this new dispersion relation
is the modification of the photon trajectory; which is the main focus in 
the present paper. (A discussion on light-lensing in compact stars is given 
by Mosquera Cuesta, de Freitas Pacheco \& Salim 2004 in a paper in preparation). 
We argue here that for extremely supercritical $B$-fields NLED effects force
photons to propagate along accelerated curves.

In case the nonlinear Lagrangean density is a function only of the scalar 
$F \equiv F_{\mu\nu} F^{\mu\nu}$, say $L(F)$, the force accelerating the 
photons is given as

\be
k_{\alpha||\nu} k^{\nu} = 
\left(4 \frac{L_{FF}}{L_{F}} F^{\mu}_{\beta}F^{\beta\nu} k_{\mu} 
k_{\nu}\right)_{| \alpha} \; , \label{acceleration}
\ee

where $k_{\nu}$ is the wavevector, $L_F$, $L_{FF}$ stands, correspondingly, 
for first and second partial derivative with respect to the invariant $F$. 
Here, and also in Eq.(\ref{k-lambda}) below, the symbols ``$_|$'' and  
``$_{||}$'' stand, respectively, for partial and covariant derivative. 
This feature allows for this force, acting along the photons path, to be 
geometrized (Novello et {al.} 2000; Novello \& Salim 2001; De Lorenci et 
al. 2002) in such a way that in an effective metric

\be
g^{\rm eff}_{\mu\nu} = g_{\mu\nu} + g^{\rm NLED}_{\mu\nu}
\ee

the photons follows geodesic paths. In such a situation, the standard
geometric procedure used in general relativity to describe the photons
can now be used upon replacing the metric of the background geometry,
whichever it is, by that of the {\it effective} metric. In this case,
the outcoming redshift proves to have now a couple of components, one
due to the gravitational field and another stemming from the magnetic
field.  As the shift in energy, and width, produced by the effective
metric ``pull" of the star on laboratory known spectral lines grows up
directly with the strength of the effective potential associated to the
effective metric, this shift has two contributions: one coming from
gravitational and another from NLED effects. In the case of hyper
magnetized stars, e. g. magnetars, both contributions may be of the same
order of magnitude. This clearly entangles imposition of constraints on
the ratio $M/R$. This difficulty can be overcome by taking in account
that the contribution of the $B$-field, which depends on both the
angles $\theta$ and $\phi$, differs from that of the gravitational
field which is fully isotropic.


Our warning (see Fig.2 !) then holds: The identification and analisys of 
spectral lines from highly magnetized NSs  must take into account the two
possible different polarizations of the received photons, in order to be 
able to discriminate between redshifts produced either gravitationally or electromagnetically. Putting this claim in perspective, we stress that if 
the characteristic redshift, or $M/R$ ratio, were to be inferred from this 
source, care should be taken since for this super strong $B$-field, 
estimated upon the identified line energy and width or via the pulsar 
spindown, such redshift becomes of the order of the gravitational one (GRS) 
expected from a canonical NS. It is, therefore, not clear whether one can 
conclusively assert something about the $M/R$ ratio of any magnetar under 
such dynamical conditions.

\subsection{The method of effective geometry}

In this paper we want to investigate the effects of nonlinearities of
very strong magnetic fields in the evolution of electromagnetic waves
here described as the surface of discontinuity of the electromagnetic
field (represented here-to-for by $F_{\mu\nu}$).  Since in the pulsar 
background there is only a magnetic field\footnote{For the present 
analysis we assume a slowly rotating NS for which the electric field 
induced by the rotating dipole is negligeable.}, then the invariant $ 
G = B_\mu E^\mu $ is not a functional in the Lagrangean. For this reason 
we will restrict our analisys  to the simple class of gauge invariant 
Lagrangians defined by

\begin{equation}
 L = L(F) \; .\label{eq1} \
\end{equation}

The surface of discontinuity\footnote{ Of course, the entire discussion 
onwards could alternatively be rephrased in terms of concepts more familiar 
to the astronomy community as that of light rays used for describing the 
propagation of electromagnetic waves in the optical geometry approximation.} 
for the electromagnetic field will be represented by $\Sigma$. We also assume 
that the field $F_{\mu\nu}$ is continuous when crossing $\Sigma$ and that its 
first derivative presents a finite discontinuity (Hadamard 1903):

\be 
\left[F_{\mu\nu}\right]_{\Sigma}=0\; ,\label{eq2} \ 
\ee

and 

\be
\left[F_{\mu\nu|\lambda}\right]_{\Sigma}=f_{\mu\nu}k_\lambda\; ,\label{eq3}
\ 
\ee 

respectively. The symbol

\be 
\left[F_{\mu\nu}\right]_{\Sigma} \equiv {\rm lim}_{\delta \rightarrow 0^+}(J|_{\Sigma+\delta}-J|_{\Sigma-\delta}) \; 
 \ee

 represents the discontinuity of the field through the surface
 $\Sigma$. The tensor $f_{\mu\nu}$ is called the discontinuity of
 the field, whilst 

 \be
 k_{\lambda} \equiv \Sigma_{|\lambda} \label{k-lambda}
 \ee

is called the propagation vector. From the least action principle we obtain 
the following field equation

 \be
 \left(L_F F^{\mu\nu}\right)_{||\mu} = 0 \; .\label{eq4} \
 \ee

Applying the Hadamard conditions (\ref{eq2}) and (\ref{eq3}) to 
the discontinuity of the field in Eq.(\ref{eq4}) we obtain

\be 
L_F f^{\mu\nu}k_\nu+2L_{FF} \xi F^{\mu\nu}k_\nu=0\; ,\label{eq5} 
\ee 

where $\xi$ is defined by

\be 
\xi\doteq F^{\alpha\delta}f_{\alpha\delta}\; .
 \ee

Both, the discontinuity conditions and the electromagnetic field tensor 
cyclic identity lead to the following dynamical relation

 \be
f_{\mu\nu}k_{\lambda}+f_{\nu\lambda}k_{\mu}+f_{\lambda\mu}k_{\nu}=0
\label{eq6} \; .
\ee 

In the particular case of a polarization such that $\xi=0$, it follows 
from Eq.(\ref{eq4}) that 

\be
f^{\mu\nu}k_\nu=0 \label{eq7} \; .
\ee

Thus, by multiplying Eq.(\ref{eq6}) by $k^\lambda$ and using the result 
of Eq.(\ref{eq7}) we obtain

 \be
f_{\mu\nu} k_{\lambda}k^{\lambda} = 0\; .
 \ee

This equation expresses that for this particular polarization the
discontinuity propagates with the metric $f_{\mu\nu}$ of the 
background space-time. For the general case, when $\xi \neq 0$, 
we multiply Eq.(\ref{eq6}) by $k_\alpha g^{\alpha\lambda} F^{\mu\nu}$ 
to obtain

\be 
\xi k_\nu k_\mu g^{\mu\nu}+2F^{\mu\nu}f_\nu^\lambda k_\lambda k_\mu = 0 \; . 
\ee

From this relation and  Eq.(\ref{eq5}) we obtain the propagation
law for the field discontinuities, in this case given as 

\be 
\left(L_F g^{\mu\nu} - 4L_{FF} F_{\alpha}^{\mu} F^{\alpha\nu} \right)
k_\mu k_\nu = 0 \; ,  \label{ef-metric}
\ee 

where

\be 
F^{\mu }_{{\;} {\;} {\;} \alpha}  F^{\alpha \nu}  = - B^2 h^{\mu \nu} 
- B^\mu B^\nu \label{F-contracted} \; .
\ee

Eq.(\ref{ef-metric}) allows to interpret the term inside the parenthesis 
multiplying $k^\mu k^\nu $ as an effective geometry 

\be 
g^{\mu\nu}_{\rm eff} = L_Fg^{\mu\nu}-4L_{FF} F^{\mu}_{\alpha} 
F^{\alpha\nu}\; . \label{eff-metric} 
\ee 

Hence, one concludes that the discontinuities will follow geodesics in this 
effective metric.

\subsection{Born-Infeld NLED}

One can start the study of the NLED effects  on the light propagation from
hypermagnetized neutron stars, with the {Born-Infeld} (B-I) Lagrangean (recall 
that a typical pulsar has no relevant electric field, i.e., $E$ is null) 

\be 
L = L(F) \; ,
\ee

with

\be 
L  = -\frac{b^2}{2} \left[ \left(1 + \frac{F}{
b^2}\right)^{1/2} - 1 \right] \; , \label{B-I-lagrangean} 
\ee

where

\be 
b = \frac{e}{R_0^2} = \frac{e}{\frac{e^4}{m_0^2 c^8}} \longrightarrow b =
\frac{m_0^2 c^8}{e^3} = 9.8 \times 10^{15}\;  {\rm e.s.u.} \; 
\ee 

In order to obtain the effective metric that decurs from the B-I Lagrangean,
one has therefore to compute the derivatives of the Lagrangean with respect 
to $F$. The first of them reads

\be 
L_{\rm F} = \frac{-1}{4 \left(1 + \frac{F}{ b^2}\right)^{1/2}
}\; , \label{B-I-derivate1} 
\ee

while its second derivative follows as

\be 
L_{\rm FF} = \frac{1}{8 b^2 \left(1 + \frac{F}{
b^2}\right)^{3/2} } \; . \label{B-I-derivate2} 
\ee

The $L(F)$ B-I Lagrangean produces, according to equation (\ref{eff-metric}), 
an {\it effective} contravariant metric given as

\be 
g^{\mu \nu}_{\rm eff}  = \frac{-1}{4 \left(1 + \frac{F}{
b^2}\right)^{1/2} } g^{\mu \nu} + \frac{B^2}{2 b^2 \left(1 +
\frac{F}{ b^2}\right)^{3/2} } \left[ h^{\mu \nu} + l^\mu l^\nu
\right] \; , 
\ee

where we define the tensor $h_{\mu\nu}$ as the metric induced in the
reference frame perpendicular to the observers; which are determined 
by the vector field $V^\mu$, and $l^\mu \equiv \frac{B^\mu}{|B^\gamma
B_\gamma|^{1/2}}$ as the unit 4-vector along the $B$-field direction.

Because the geodesic equation of the discontinuity (that defines
the effective metric) is conformal invariant, one can multiply 
this last equation by the conformal factor

\be 
{-4} \left(1 + \frac{F}{ b^2}\right)^{3/2} 
\ee

to obtain

\be 
g^{\mu \nu}_{\rm eff}  = \left(1 + \frac{F}{ b^2}\right)
g^{\mu \nu} - \frac{2 B^2}{b^2} \left[ h^{\mu \nu} + l^\mu l^\nu
\right] \; . 
\ee

By noting that

\be 
F =  F_{\mu \nu} F^{\mu \nu} = - 2 (E^2 - B^2) \; ,
\ee

and recalling that in the case of a canonical pulsar $E = 0$,  
then $F = 2 B^2$. Therefore, the effective metric reads

\be 
g^{\mu \nu}_{\rm eff}  = (1 + \frac{2 B^2}{ b^2}) g^{\mu \nu}
-\frac{2 B^2}{ b^2} \left[ h^{\mu \nu} + l^\mu l^\nu \right] \; ,
\ee

or equivalently

\be 
g^{\mu \nu}_{\rm eff}  =  g^{\mu \nu}  + \frac{2 B^2}{ b^2}
V^\mu V^\nu - \frac{2 B^2}{ b^2} l^\mu l^\nu \; . 
\ee

As one can check, this effective metric is a functional of the
background metric $g^{\mu \nu}$, the 4-vector velocity field of 
the inertial observers $V^\nu$, and the spatial configuration 
(orientation $l^\mu$) of the $B$-field.

Because the concept of gravitational redshift (GRS) is associated with 
the covariant form of the background metric, one needs to find the inverse 
of the effective metric $g^{\mu \nu}_{\rm eff}$ given above. With the 
definition of the inverse metric

\be 
g^{\mu \nu}_{\rm eff} g_{\nu \alpha}^{\rm eff} = \delta^{\mu}_{{\;} {\;} 
\alpha} \; , 
\ee

one obtains the covariant form of the effective metric as 

\be 
g_{\mu \nu}^{\rm eff} =  g_{\mu \nu} - \frac{2B^2/b^2}{
(2B^2/b^2 + 1) } V_\mu V_\nu  + \frac{2B^2/b^2}{ (2B^2/b^2 + 1) }
l_\mu l_\nu \; . \label{effective-metric} 
\ee


In order to openly write the covariant time-time effective metric
component, one can start by figuring out that a) both the emitter 
and observer are in inertial frames, then $V^\mu = {\delta^\mu_0}/
(g_{00})^{1/2} $, and b) the magnetic field is a pure radial field. 
In this case, one can write $B^\mu = B^\mu(r)$, where $B^\mu \equiv
B~l^\mu = B~l^r$, which implies that the $l^\mu$ time, polar and
azimuthal vector components become $l^t = 0$, $l^\theta = 0$ and
$l^\phi = 0$. By using all these assumptions in Eq.(\ref{eff-metric})
one arrives to the time-time effective metric component (in the 
{\it appendix} we derive the corresponding $g_{r r}$ metric 
component, which is also interesting to reckon with)

\be 
g_{t t}^{\rm eff} = g_{t t} - \frac{2B^2/b^2}{ (2B^2/b^2 +
1) }~~g_{t t} \; , 
\ee

or similarly

\be
g_{t t}^{\rm eff} = \left[\frac{1}{ 2B^2/b^2 + 1 }\right]~ g_{t t}\; .
\label{tt-component}
\ee

\begin{figure}
\centering
\resizebox{7.4cm}{!}{
\includegraphics{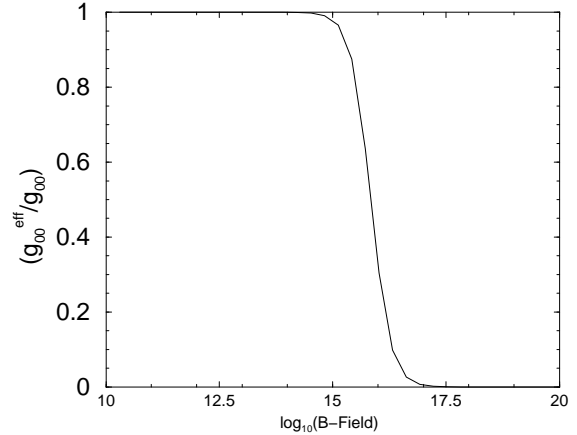} } 
\caption{Ratio between the effective time-time metric components
$g_{00}^{\rm eff}$ and the background $g_{00}$  as a function of the
magnetic field strength $B$-Field.  Notice that the ratio goes to zero
as larger the $B$-field grows, which means that the {\it effective}
(gravitational + NLED) redshift tends to infinity while the standard
one (GRS) remains constant for a given $M/R$, as it independs on $B$.}
\end{figure}

This effective metric corresponds to the result already derived in our
previous paper (Mosquera Cuesta \& Salim 2003;2004). We stress,
meanwhile, that our former result was obtained by using the approximate
Lagrangean

\be
L(F) = - \frac{1}{4} F + \frac{\mu}{4} \left(F^2 + \frac{7}{4} G^2\right) 
\; , \label{E-H-LAGRANGEAN}
\ee

where $\mu = \frac{2 \alpha^2}{45} \frac{(\hbar/m c)^3}{m c^2}$, with
$\alpha = \frac{e^2}{4 \pi \hbar c}$, $G \equiv F_{\mu\nu} F^{\ast \;
\mu\nu}$, $F^{\ast \; \mu\nu} \equiv \frac{1}{2} \eta^{\mu \nu \alpha
\beta} F_{\alpha \beta}$, and $F$ was defined above. This Lagrangean is
built up on the first two terms of the infinite series expansion
associated with the Heisenberg-Euler (1936) Lagrangean, which proved to
be valid for $B$-field strengths near the quantum electrodynamics
critical field $B\sim 10^{13.5}$~G. In the present paper we overrun
that limit.  In fact, from Eq.(\ref{tt-component}) is straightforward
to verify that the ratio $g_{00}^{\rm eff}/g_{00} \leq 1$, and goes all 
the way down to zero as the $B$-field attains higher strength values. 
This means that the {\it effective} surface redshift grows unbounded as 
$B$ becomes larger and larger. Both results are displayed in Figs.1 and 2.
Fig.2 also confirm that for canonical NSs the gravitational (GRS) redshift
remains constant even for fields larger then the Schafroth QED limit.

\begin{figure}
\centering
\resizebox{7.4cm}{!}{
\includegraphics{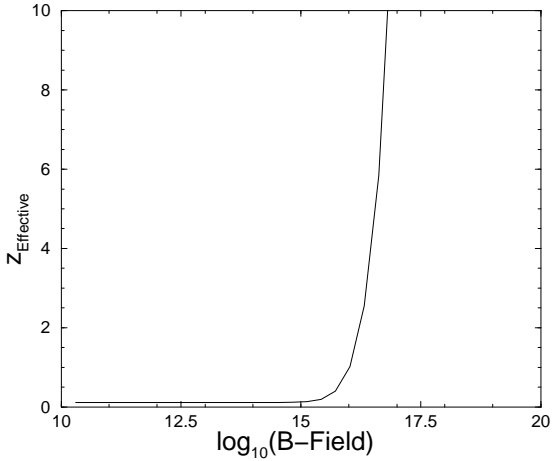} } 
\caption{The plot shows the variation, for a given constant $M/R$ ratio, of the 
effective redshift (which in turns depends on the effective time-time metric 
component $g_{00}^{\rm eff}$) as a function of the $B$-Field strength. Notice 
that it tends to infinity as larger the $B$-field grows, while the standard 
one (GRS) takes on a fixed value $z\leq 1$.  }
\end{figure}

\section{Discussion and Conclusion}

In a very interesting couple of {\it Letters} by Ibrahim et {al.}
(2002;2003) was reported the discovery of cyclotron line
resonance features in the source SGR 1806-20, said to be a candidate to
a magnetar by Kouveliotou {et al.} (1998).  The 5.0 keV feature
discovered with {\it Rossi XTE} is strong, with an equivalent width of
$\sim 500$~eV and a narrow width of less than 0.4~eV (Ibrahim et {al.}
2002;2003). When these features are interpreted in the context of
accretion models one arrives to a $M/R > 0.3$ M$_\odot$
km$^{-1}$, which is inconsistent with NSs, or that requires a low
$(5-7) \times 10^{11}$~G magnetic field, which is said not to
correspond to any SGRs (Ibrahim et {al.} 2003). In the magnetar
scenario, meanwhile, the features are plausibly explanined as being
ion-cyclotron resonances in an ultra strong $B$-field $B_{\rm sc} \sim
10^{15}$~G (see Eq.(\ref{line})).  The spectral line is said to be
consistent with a proton-cyclotron fundamental state whose energy and
width are close to model predictions (Ibrahim et { al.} 2003).
According to Ibrahim et {al.} (2003), the confirmation of this findings
would allow to estimate the gravitational redshift (the GRS), mass ($M$)
and radius ($R$) of the quoted ``magnetar'': SGR 1806-20.

Although the quoted spectral line in Ibrahim et {al.} (2002;2003) is
found to be a cyclotron resonance produced by protons in that high
field, we alert on the possibility that it could also be due to NLED
effects in the same super strong magnetic field of SGR 1806-20, as
suggested by Eq.(\ref{tt-component}).  If this were the case, no
conclusive assertion on the $M/R$ ratio of the compact star glowing in
SGR 1806-20 could be consistently made, since a NLED effect might well
be emulating the standard gravitational one associated with the pulsar
surface redshift.

In summary, because we started with a more general and exact
Lagrangean, as that of Born \& Infeld (1934), we can assert that the
result here derived is inherently generic to any kind of nonlinear
theory describing the electromagnetic interaction, and thus is
universal in nature. Thence, absorption or emission lines from
magnetars, if these stellar objects do exist in nature (see P\'erez
Mart\'{\i}nez et al. 2003 for arguments contending their formation),
cannot be saftely used as an unbiased source of information regarding
the fundamental parameters of a NS pulsar such as its $M$, $R$ or EOS.

\section*{Appendix}

An additional outcome of the above procedure regards the modifications 
to the radial component of the background metric. For a pure radial
$B$-field the 4-vector unit $l^\mu$ can be written as

\be 
l_\mu \equiv \frac{\delta_\mu^r}{\sqrt{- g^{r r} } } = \sqrt{- g_{r r} } ~~ 
\delta^r_\mu \; , 
\ee

which renders  

\be 
l^r = \frac{\delta^\mu_r}{\sqrt{- g_{r r} } } \; , 
\ee

and consequently

\be 
l^\mu l_\mu = -1 \longrightarrow \longrightarrow  l_r l_r g^{r r} = - 1 \; .
\ee

Therefore, the third term in Eq.(\ref{effective-metric}) reduces to 

\be 
B^\mu B^\nu g_{\mu \nu} = - B^2 l^r l^r g_{r r} \; . 
\ee

In this way, one can verify that such a radial-radial effective metric 
component is given by the relation

\be 
g_{r r}^{\rm eff} = g_{r r} - \frac{2B^2/b^2}{ (2B^2/b^2 +
1) }~  g_{r r} = \left(1 - \frac{2B^2/b^2}{ [2B^2/b^2 + 1] }
\right)~g_{r r}  
\ee

or equivalently

\be 
g_{r r}^{\rm eff}  =  \left[\frac{1 }{ 1 + 2B^2/b^2 }\right]~ 
g_{r r} \; . 
\ee

Some astrophysical consequences of this fundamental change in the radial 
component of the background metric will be explored in a forthcoming paper.

\section*{Acknowledgements} 

We are grateful to the anonymous referee for all the suggestions and
criticisms that helped us to improve this manuscript.  J. M. Salim
acknowledges Conselho Nacional de Desenvolvimento Cient\'{\i}fico e
Tecnol\'ogico (CNPq/Brazil) for the Grant No. 302334/2002-5.  HJMC
thanks Funda\c c\~{a}o de Amparo \`a Pesquisa do Estado de Rio de
Janeiro (FAPERJ/ Brazil) for the Grant-in-Aid 151.684/2002.

\end{document}